\documentclass[acmsmall, screen, authorversion]{acmart}

\usepackage{amsmath,amsfonts}
\usepackage[ruled,linesnumbered]{algorithm2e}
\usepackage{tikz}
\usetikzlibrary{positioning}
\usepackage{mathtools}
\usepackage{mathpartir}
\usepackage{listings}
\usepackage{pifont}
\usepackage{caption}
\usepackage{subcaption}
\usepackage{tcolorbox}
\usepackage{mathdots}

\definecolor{highlight}{RGB}{180,30,30}

\definecolor{codeborder}{RGB}{240, 240, 240}
\NewTotalTCBox{\rs}{ s v }
{verbatim,colback=white,colframe=codeborder,boxsep=0mm,left=2pt,right=2pt,top=2pt,bottom=2pt}
{\lstinline^#2^}

\newcounter{typerule}

\newcommand{\ir}[4]{%
    \def\thetyperule{#1}%
    \refstepcounter{typerule}%
    \label{rule:#4}%
    \inferrule[#1]{#2}{#3}
}
\newcommand{\lrangle}[1]{\langle#1\rangle}

\AtBeginDocument{%
  }

\acmISBN{978-1-4503-XXXX-X/18/06}

\begin{document}

\title{Possible Value Analysis based on Symbolic Lattice}

\author{Qi Zhan}
\email{qizhan@zju.edu.cn}
\orcid{1234-5678-9012}
\affiliation{%
    \institution{Zhejiang University}
    \city{Hangzhou}
    \country{China}
}

\begin{abstract}
    We propose a new static program analysis called program behavior analysis. 
    The analysis aims to calculate possible symbolic expressions for every variable at each program point. 
    We design a new lattice, transfer function, and widening operator to accommodate the analysis. 
    Furthermore, we extend the analysis to interprocedural.
\end{abstract}


\keywords{program analysis}

\maketitle

\section{Introduction}

Static program analysis is a well-studied field~\cite{ai}, which aims to reason about the behavior of computer programs without actually running them. Static program analysis is useful not only in optimizing compilers but also provide static bug or vulnerability detection tools~\cite{vsa, arzt2014flowdroid}.

In this paper, we propose a new static analysis technique, called program behavior analysis.
At a high level, we combine symbolic representation~\cite{symbolicexecution} and abstract interpretation~\cite{ai} to calculate possible symbolic values for every variable and memory cell at each program point.
With symbolic representation, the whole value flow is tracked by the initial function parameters and global variables and their corresponding memory structure;
with abstract interpretation, possible values are represented as an element of a lattice, where the top element $\top$ represents any value. The definition of the lattice is interleaved with the symbolic expression, which is capable of representing any symbolic value.

As the lattice is infinite height, we propose a special widening technique to approximate the transfer function. In addition, the values from different control flows are joined together naturally based on the join of lattice.
By running the classical worklist algorithm in static analysis, we are able to get precise function behavior from fixpoint results.
The semantic level of the code diff can be obtained by comparing the behavior of the function.
Furthermore, we believe that the semantic model can also be used to describe the behavior of the whole program rather than limited to code diff, so we extend the analysis to interprocedural.
We discuss their potential usage in future work.

The main contribution of this paper is a new type program behavior analysis based on symbolic representation and we extend it to interprocedural.

\section{Possible Value Analysis}\label{sec:approach}

\subsection{Preliminary}\label{sec:preliminary}

In this section, we present some basic techniques and terminology widely used in academic research for program analysis that constitute the foundation of our work.

\textbf{Abstract Interpretation}~\cite{ai} is a well-established foundation for static program analysis.
In abstract interpretation, we abstract the value as an element in abstract domain, and each program statement is given an interpretation over abstract domain.
This framework guarantees that the results of analysis overapproximate the concrete behavior of the program, which is called ``soundness''.

\textbf{Pointer Analysis} is a particular program analysis technique, that aims to compute the points-to information of the program, which is used in compiler optimization, program verification, and program understanding~\cite{jensen2009type,arzt2014flowdroid,chandra2009snugglebug}.
A typical flow-sensitive Andersen's algorithm is shown in Fig.~\ref{fig:pointer-analysis}.
IN$_v^a$ and OUT$_v^a$ represent the input and output points-to set of the variable $a$ at the program point $v$, respectively.
For store statement, we separate the weak update and strong update~\cite{strongupdate}.
In the case that the input points-to-set of the left-hand side variable is singleton, we can update the memory value directly.
When not, we must update the points-to relation by unioning the original set.

\begin{figure}[H]
    \begin{tabular}{|l|ll|}
        \hline
        Statement  & Constrains                                                                              &                       \\
        \hline
        $a=\&b$    & $\text{OUT}_v^a = \{b\}$                                                                &                       \\
        $a=b$      & $\text{OUT}_v^a = \text{IN}_v^b$                                                        &                       \\
        $a=\ast b$ & $\text{OUT}_v^a = \bigcup_{b' \in \text{IN}_v^b} \text{IN}_v^{b'}$                      &                       \\
        $\ast a=b$ & $\forall a' \in \text{IN}_v^a, \text{OUT}_v^{a'} = \text{IN}_v^b$                       & $|\text{IN}_v^a| = 1$ \\
                   & $\forall a' \in \text{IN}_v^a, \text{OUT}_v^{a'} = \text{IN}_v^{a'} \cup \text{IN}_v^b$ & $|\text{IN}_v^a| > 1$ \\
        \hline
    \end{tabular}
    \caption{A typical flow sensitive Andersen's algorithm}
    \label{fig:pointer-analysis}
\end{figure}

\textbf{Symbolic Execution}~\cite{symbolicexecution} is a way of executing a program abstractly.
The main idea is to treat concrete inputs as symbols and return symbolic expressions expressed in input values.
An abstract execution can cover multiple possible inputs of the program that share a particular execution path through the code.
Our approach relies on the idea that using symbolic execution to capture the possible value of the program,
while we do not consider the solution of symbolic constraints to gain concrete inputs.

\subsection{Overview}\label{sec:overview}
\SetKwComment{Comment}{/* }{ */}
\begin{algorithm}
    \KwIn{IN and OUT state of every basic block.}
    \ForEach{basic block B} {
        OUT[B] = $\emptyset$\;
    }
    W $\gets$ all basic blocks\;
    \While{W is not empty}{
    Pick a basic block B from W\;
    old = OUT[B]\;
    IN[B] = $\bigcup_{P~{\text{is a predecessor of B}}}$ OUT[P] \Comment*[r]{ Section~\ref{sec:join}}
    OUT[B] = \textit{transfer}(IN[B]) \Comment*[r]{ Section~\ref{sec:intraprocedural}}
    OUT[B] = IN[B] $\nabla$ OUT[B]\Comment*[r]{ Section~\ref{sec:widening}}
    \If{old $\ne$ OUT[B]} {
        Add all successors of B to W\;
    }
    }
    \caption{Worklist Algorithm}
    \label{alg:worklist}
\end{algorithm}

Possible value analysis computes the abstract value of each variable and memory cell at each program point.
The overall framework of possible value analysis is illustrated in Algorithm~\ref{alg:worklist}.
It is an abstract interpretation to find a safe approximation for the set of symbolic expressions that each data object holds at each program point.
The main procedure is same as common data flow analysis, i.e. we use worklist algorithm described in Alg.~\ref{alg:worklist} to compute the least fix point.
We assume the control flow graph of the function is given, and the analysis is performed on the control flow graph.
Worklist Algorithm is a typical technique used in the monotone framework of data flow analysis~\cite{kam1977monotone}.
It maintains IN and OUT sets for each basic block, and a work list iteratively updates the data flow information of the basic blocks.
When OUT state of a basic block is updated, all successors of the basic block are added to the worklist.
The algorithm terminates until the fixpoint is reached, i.e. worklist is empty.
As the algorithm can be used to solve data flow analysis constraints, and possible value analysis belongs to the category,
all we need to do is to instantiate abstract domain and transfer function in the algorithm:

\begin{enumerate}
    \item Abstract domains. In Section~\ref{sec:formalization}, we define the lattice used in the analysis, as well as the widening technique~\ref{sec:widening}, and the abstract domains are defined in Fig.~\ref{fig:abstract-domains}.
    \item Transfer function. It interprets every program statement in the abstract domain, we discuss it in Section~\ref{sec:intraprocedural}.
\end{enumerate}

\subsection{Formalization}\label{sec:formalization}

\subsubsection{Language.}
We formalize our approach using a simple SSA form~\cite{ssa} call-by-value language in Fig.~\ref{fig:language}, following the previous paper~\cite{valueflow, valueflow2} in the program analysis community.
The language's syntax and semantics are straightforward and contain the basic features of a programming language.
Compared to~\cite{valueflow, valueflow2}, we add a global variable declaration to the language, which is necessary to depict the behavior related to the global variable.
In addition, we assume that the last statement of a function is always a return statement to facilitate the discussion and inference rule.

\begin{figure}
    \begin{subfigure}[t]{0.45\linewidth}
        \centering
    \begin{align*}
        \text{Program}~\textit{P}   & \coloneq (F \mid G)^+                                 \\
        \text{Function}~\textit{F}  & \coloneq f(v_1,\dots,v_n)\{B*; \}                           \\
        \text{Block}~\textit{B}     & \coloneq S                                                  \\
        \text{Statement}~\textit{S} & \coloneq v_1 = v_2 \mid v_1 = \&v_2                         \\
                                    & \mid v_1 = \ast v_2 \mid \ast v_1 = v_2                     \\
                                    & \mid v_1 = v_2~\text{binop}~v_3 \mid v_1 = \text{unop } v_2 \\
                                    & \mid \text{if} (v)~\text{then}~\text{goto}~B                \\
                                    & \mid v = \text{call}~f(v_1,\dots,v_n) \mid \text{return}~v  \\
                                    & \mid v = \phi(v_1, \dots, v_n)                                  \\
                                    & \mid s_1; s_2                                               \\
        \text{binop }               & \coloneq + \mid - \mid \ast \mid /  \mid \cdots             \\
        \text{unop }                & \coloneq - \mid \neg \mid  \cdots \\
        \text{Global } &g \in G
    \end{align*}
    \caption{Syntax of the language}
    \label{fig:language}
    \end{subfigure}
    \begin{subfigure}[t]{0.45\linewidth}
        \centering
        \begin{align*}
            \text{Expression}~\textit{e}      & \coloneq \text{primitive}~p                     \\
                                              & \mid \text{function}(l_1,\dots,l_n)             \\
                                              & \mid \text{binop}(l_1, l_2)                     \\
                                              & \mid \text{unop}(l)                             \\
                                              & \mid \phi(l_1,\dots,l_n)                         \\
            \text{Primitive}~\textit{p}       & \coloneq \text{const}(i)                        \\
                                              & \mid \text{arg}(i)                              \\
                                              & \mid \text{mem}(o)                              \\
                                              & \mid \text{global}(o)                           \\
            \text{binop }                     & \coloneq + \mid - \mid \ast \mid /  \mid \cdots \\
            \text{unop }                      & \coloneq - \mid \neg \mid  \cdots               \\
            \\
            \text{Lattice Element}~\textit{l} & \coloneq \top  \mid \bot  \mid  \text{Expression}(e) 
        \end{align*}
        \caption{Syntax of symbolic expression and lattice}
        \label{fig:lattice}
    \end{subfigure}
    \caption{Syntax of language and lattice}
\end{figure}

\subsubsection{Lattice Design.}\label{sec:lattice}

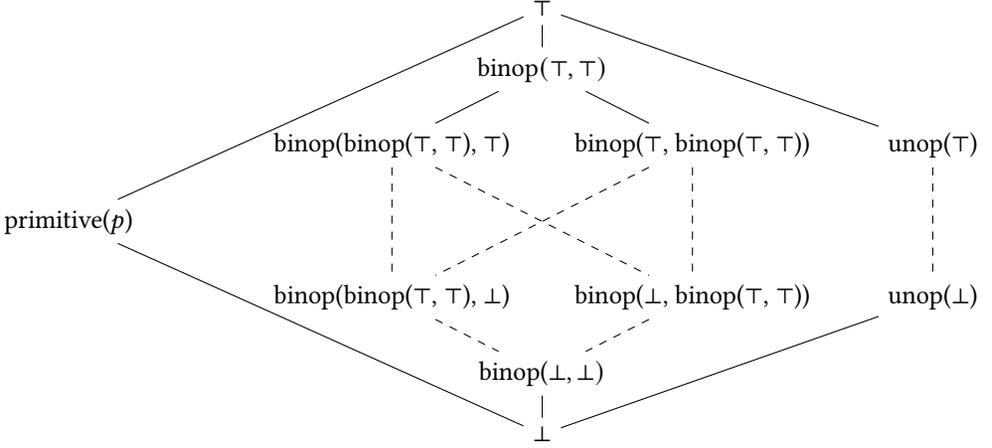
\begin{figure}
        \centering
        \begin{tikzpicture}
            \node (top) at (0,5.8) {$\top$};
            \node (bintop) at (0,5) {binop$(\top, \top)$};
            \node (binltop) at (-2, 4) {binop(binop($\top$, $\top$), $\top$)};
            \node (binlbot) at (-2, 2) {binop(binop($\top$, $\top$), $\bot$)};
            \node (binrtop) at (2, 4) {binop($\top$, binop($\top$, $\top$))};
            \node (binrbot) at (2, 2) {binop($\bot$, binop($\top$, $\top$))};
            \node (binbot) at (0, 1) {binop($\bot, \bot)$};
            \node (p) at (-6.3, 3) {primitive($p$)};
            \node (utop) at (5.2, 4) {unop($\top$)};
            \node (ubot) at (5.2, 2) {unop($\bot$)};
            \node (bot) at (0,0.2) {$\bot$};

            \draw (top) -- (p) -- (bot);
            \draw (top) -- (utop);
            \draw [dashed] (utop) -- (ubot);
            \draw (ubot) -- (bot);
            \draw (bintop) -- (top);
            \draw (binbot) -- (bot);
            \draw (bintop) -- (binltop);
            \draw (bintop) -- (binrtop);
            \draw [dashed] (binltop) -- (binlbot) -- (binbot);
            \draw [dashed] (binrtop) -- (binrbot) -- (binbot);
            \draw [dashed] (binltop) -- (binrbot);
            \draw [dashed] (binrtop) -- (binlbot);
        \end{tikzpicture}
        \caption{Illustration of a subset of the lattice. For simplicity, we only consider the primitive $p$ and binary expressions and unop expressions. Function and phi expression can be added in a similar way.
        As the lattice is infinite height, we have to omit some elements, which are represented as a dashed line. 
        }
        \label{fig:lattice-illustration}
\end{figure}

As with every static analysis method, we design a lattice to abstract the possible values of the program, as shown in Fig.~\ref{fig:lattice}.
In our work, the definition of lattice $\mathcal{L}$ is unusual, as it recursively interleaved with the definition of the expression $e$.

\textbf{Lattice} $l$ is the abstract value of possible symbolic expressions.
          $\top$ and $\bot$ are the maximal and minimal elements of the lattice, respectively.
          For example, $f(\top, \text{binop}(\top, \text{const}(1)))$ is a valid lattice element. It can be understood as a specific function $f$, where the first argument is any possible value, and the second argument is a binary operator of any possible value and const. The information is more precise than $f(\top, \top)$ since we can restrict the possible constructions for the second argument.

\textbf{Expression} $e$ can be a primitive value, a function call, a binary operation, a phi node, or a unary operation,
          which corresponds to the concrete syntax of the language in Fig.~\ref{fig:language}.
          Memory $o$ here is a memory location consisting of a base address and an offset, where the

The partial order $\sqsubseteq$ and the join operator of set $\mathcal{L}$ are defined in Fig.~\ref{fig:partial-order} and Table~\ref{tab:join} recursively.
For the same constructor, the partial order and join are defined as the partial order and join of the corresponding fields, respectively.
It is easy to check that the order set equipped with the above operators is \textit{indeed} a complete lattice.
Every subset has a least upper bound and a greatest lower bound, and the join and meet operators are idempotent, commutative, and associative.
In the discussion below, we use $\mathcal{L} = \lrangle{L, \sqsubseteq, \sqcup, \bot, \top}$ to represent the lattice.

It seems that the lattice is too simple and not expressive enough, as it consists of only one non-trivial element $e$.
The magic lies in the fact that the expression can contain lattice elements. An expression $e$ can contain lattice elements, and a lattice element can contain expression $e$, the lattice can be any symbolic expression.
The strong expressiveness of the lattice makes it possible to not miss any single value of the program.
On the other hand, the height of the lattice is infinite, which does not satisfy the termination requirement of the worklist algorithm directly.
Fig.~\ref{fig:lattice-illustration} illustrates a subset of the lattice, which only consists of primitive value and binary operator.

\begin{figure}
    \begin{align*}
        \bot                            \sqsubseteq l                                                                                                     \\
        l                               \sqsubseteq \top                                                                                                  \\
        p = p'                                                 & \Rightarrow \text{primitive}(p) \sqsubseteq \text{primitive}(p')                         \\
        l_i \sqsubseteq l_i' \text{ for } i = 1,\dots,n        & \Rightarrow \text{function}(l_1,\dots,l_n)  \sqsubseteq \text{function}(l_1',\dots,l_n') \\
        l_i \sqsubseteq l_i' \text{ for } i = 1,\dots,n        & \Rightarrow \phi(l_1,\dots,l_n) \sqsubseteq \phi(l_1',\dots,l_n')                        \\
        l_1 \sqsubseteq l_1' \text{ and } l_2 \sqsubseteq l_2' & \Rightarrow \text{binop}(l_1,l_2) \sqsubseteq \text{binop}(l_1',l_2')                    \\
        l \sqsubseteq l'                                       & \Rightarrow \text{unop}(l)                  \sqsubseteq \text{unop}(l')
    \end{align*}
    \caption{Partial order of lattice}
    \label{fig:partial-order}
\end{figure}

\begin{table}
    \caption{Join of lattice element}
    \begin{tabular}{lll}
        \toprule
        \textbf{left}                 & \textbf{right}              & \textbf{join $\sqcup$}                          \\ \midrule
        $\bot$                        & $l$                         & $l$                                             \\
        $l$                           & $\bot$                      & $l$                                             \\
        primitive($p$)                & primitive($p'$)             & primitive($p$), if $p=p'$; $\top$, otherwise    \\
        function($l_1,\dots,l_n$)     & function($l_1',\dots,l_n'$) & function($l_1\sqcup l_1',\dots,l_n\sqcup l_n'$) \\
        $\phi(l_1,\dots,l_n)$         & $\phi(l_1',\dots,l_n')$     & $\phi(l_1\sqcup l_1',\dots,l_n\sqcup l_n')$     \\
        binop($l_1,l_2$)              & binop($l_1',l_2'$)          & binop($l_1\sqcup l_1',l_2\sqcup l_2'$)          \\
        unop($l$)                     & unop($l'$)                  & unop($l\sqcup l'$)                              \\
        \multicolumn{2}{c}{otherwise} & $\top$                                                                        \\ \bottomrule
    \end{tabular}

    meet $\sqcap$ is defined similarly, we omit it here.
    \label{tab:join}
\end{table}

\subsubsection{Widening}\label{sec:widening}
Since the height of the lattice is not limited, the worklist algorithm may not terminate.
For example, consider a very simple add statement $ a = a + 1 $ in a loop, a will be evaluated as (+ a 1), (+ (+ a 1) 1), and so on.
Widening technique is introduced by~\cite{ai} to overcome this limitation.
In this section, we propose a series of simple widening strategies for our lattice.

The intuition for widening comes from the recursive definition of lattice.
As shown in Fig.~\ref{fig:lattice-illustration}, the lattice can be seen as a layer hierarchy.
$\top,\bot,\text{primitive}(p)$ consist the first layer of the lattice.
binop$(\top, \top)$, binop$(\top, \text{primitive})$ consit the second layer and so on.
It is easy to define a depth function $\mathcal{D}$ of an element in lattice:

\begin{align*}
    \mathcal{D}(\top)                            = 0, \mathcal{D}(\bot) & = 0, \mathcal{D}(\text{primitive}(p)) = 0           \\
    \mathcal{D}(\text{function}(l_1,\dots,l_n))                         & = 1 + \max(\mathcal{D}(l_1),\dots,\mathcal{D}(l_n)) \\
    \mathcal{D}(\text{binop}(l_1,l_2))                                  & = 1 + \max(\mathcal{D}(l_1),\mathcal{D}(l_2))       \\
    \mathcal{D}(\text{unop}(l))                                         & = 1 + \mathcal{D}(l)                                \\
    \mathcal{D}(\phi(l_1,\dots,l_n))                                    & = 1 + \max(\mathcal{D}(l_1),\dots,\mathcal{D}(l_n))
\end{align*}

Based on the depth function, we can define a series of widening operators $\nabla_i(x, f(x))$, where we only consider $f(x)$ in the simple widening strategy.
Only the element with a depth less than or equal to $i$ is kept unchanged, and the element with a depth greater than $i$ is widened to the corresponding element in $i$-th layer.
For examle, $\nabla_0(x, \text{binop}(\top, \top)) = \top.$
The height of the sublattice $L_i$ is limited by $i$, and the worklist algorithm will always terminate in $L_i$.

\subsubsection{Abstract Domains.}\label{sec:abstractdomain}

Based on the definition of lattice, abstract domains can be defined as usual.
The symbols and abstract domains are listed in Fig.~\ref{fig:abstract-domains}, and introduced as follows:
\begin{itemize}
    \item Expression $e$ is a value expression in the lattice $\mathcal{L}$, variables $v$ are used to represent the local variables,
          and memory cell $o$ is used to represent the memory locations.
    \item The local variable context $\Sigma$ is a mapping from variable names to value expressions.
    \item The memory $m$ is a mapping from memory locations to value expressions.
          We regard global variables as a part of memory locations.
    \item The points-to relation $\Delta$ is a mapping of memory locations to the set of memory locations,
          representing the possible locations to which a memory cell can point.
\end{itemize}

\begin{figure}
    $$
        \begin{aligned}
             & \begin{array}{ll}
                   \text{Expression}~e \in E                     & \text{Object}~o \in \mathcal{O} \quad\quad \text{Lattice}~l \in \mathcal{L} \\
                   \text{Variables}~v \in \mathcal{V}            & \text{Local context}~\Sigma: \mathcal{V} \to \mathcal{L}                    \\
                   \text{Memory}~m : \mathcal{O} \to \mathcal{L} & \text{Points-to relation}~\Delta : \mathcal{O} \to \mathcal{P}(\mathcal{O}) \\
               \end{array}
        \end{aligned}
    $$
    \caption{Abstract domains}
    \label{fig:abstract-domains}
\end{figure}

\subsubsection{Join Operator} \label{sec:join}
$\sqcup$ is originally used to join two elements in lattice, and is also overloaded to merge context environments for convenience.
Join of lattice element is the definition of the lattice;
Join of local variable context and memory can be reduced to the join of lattice element;
Join of points-to relation is the union of the two sets.
$\sqcup_L$ in left is the join operator of the lattice $\mathcal{L}$.
we assume that an element not in the domain of the mapping is $\bot$ or an empty set.

\begin{align*}
    \centering
    \text{Expression}(e_1)     \sqcup \text{Expression}(e_2)     & = \text{Expression}(e_1 \sqcup_L e_2)                                             \\
    \text{Local}(\Sigma_1)     \sqcup \text{Local}(\Sigma_2)     & = \text{Local}(\Sigma), \text{where}~\Sigma(v) = \Sigma_1(v) \sqcup_L \Sigma_2(v) \\
    \text{Memory}(m_1)         \sqcup \text{Memory}(m_2)         & = \text{Memory}(m), \text{where}~m(o) = m_1(o) \sqcup_L m_2(o)                    \\
    \text{Points-to}(\Delta_1) \sqcup \text{Points-to}(\Delta_2) & = \text{Points-to}(\Delta), \text{where}~\Delta(o) = \Delta_1(o) \cup \Delta_2(o)
\end{align*}

\subsection{Transfer Function}\label{sec:intraprocedural}

\begin{figure}[!t]
    \begin{subfigure}[h]{\linewidth}

        \begin{mathpar}

            \ir{S-BinOp}
            {\Sigma; m; \Delta; v_1 \Rightarrow e_1 \\ \Sigma; m; \Delta; v_2 \Rightarrow e_2}
            {\Sigma; m; \Delta; v = v_1~\text{binop}~v_2 \Downarrow \Sigma[v \mapsto e_1~\text{binop}~e_2]; m; \Delta}
            {binop}

            \ir{S-UnOp}
            {\Sigma; m; \Delta; v \Rightarrow e}
            {\Sigma; m; \Delta; v = \text{unop}~v \Downarrow \Sigma[v \mapsto \text{unop}~e]; m; \Delta}
            {unop}

            \ir{S-Seq}
            {\Sigma; m; \Delta; s_1 \Downarrow \Sigma' ; m'; \Delta'; \\ \Sigma'; m'; \Delta';  s_2 \Downarrow \Sigma''; m''; \Delta''}
            {\Sigma; m; \Delta; s_1;s_2 \Downarrow \Sigma''; m''; \Delta''}
            {seq}

            \ir{S-Call}
            {\Sigma; m; \Delta; v_i \Rightarrow e_i, \forall i \in [1,n]}
            {\Sigma; m; \Delta; v = f(v_1,\dots,v_n) \Downarrow \Sigma[v \mapsto f(e_1,\dots,e_n)]; m; \Delta}
            {call} (\textsc{intra})

            \ir{S-Return}
            {\Sigma; m; \Delta; v \Rightarrow e}
            { \Sigma; m; \Delta; \text{return}~v \Downarrow \Sigma; m; \Delta}
            {return} (\textsc{intra})

            \ir{S-Phi}
            {\Sigma; m; \Delta; v_i \Rightarrow e_i, \forall i \in [1,n]}
            {\Sigma; m; \Delta; v = \phi(v_1, \dots, v_n) \Downarrow \Sigma[v \mapsto \phi(e_1, \dots, e_n)]; m; \Delta}
            {phi}

            \ir{S-Assign}
            {\Sigma; m; \Delta; v_2 \Rightarrow e \\ \Delta; v_1 = v_2 \Downarrow_{pa} \Delta'}
            {\Sigma; m; \Delta; v_1 = v_2 \Downarrow \Sigma[v_1 \mapsto e]; m; \Delta'}
            {assign}

            \ir{S-Ref}
            {\Delta;v_1=\& v_2 \Downarrow_{pa} \Delta'}
            {\Sigma; m; \Delta; v_1 = \&v_2 \Downarrow \Sigma; m; \Delta'}
            {ref}

            \ir{S-Load}
            {e = \bigsqcup{m[o_i]}, \forall o_i \in \Delta(v_2) \\ \Delta; v_1 = * v_2 \Downarrow_{pa} \Delta'}
            {\Sigma; m; \Delta; v_1 = *v_2 \Downarrow \Sigma[v_1 \mapsto e]; m; \Delta'}
            {load}

            \ir{S-Store}
            {\Sigma; m; \Delta; v_2 \Rightarrow e \\ \forall o_i \in \Delta(v_1) \\ \Delta; *v_1 = v_2 \Downarrow_{pa} \Delta'}
            {\Sigma; m; \Delta; *v_1 = v_2 \Downarrow \Sigma; m[o_i \mapsto m[o_i] \sqcup e]; \Delta'}
            {store}

        \end{mathpar}
        \caption{Intraprocedural transfer function rules}
        \label{fig:transfer-function}
    \end{subfigure}
    \begin{subfigure}{\linewidth}

        \ir{S-Call}
        {\Sigma; m; \Delta; v_i \Rightarrow e_i, \forall i \in [1,n] \\
            \Sigma[v_1\mapsto e_1,\dots,v_n\mapsto e_n]; m; \Delta; s \Downarrow \Sigma'; m'; \Delta' \\
            \Sigma'; m'; \Delta'; v \Rightarrow e }
        {\Sigma; m; \Delta; w = f(v_1,\dots,v_n)\{{s;~\text{return}~v}\} \Downarrow \Sigma[w\mapsto e]; m'; \Delta'} {call2} (\textsc{inter})

        \caption{Interprocedural transfer function rules}
        \label{fig:transfer-function-inter}
    \end{subfigure}
    \caption{Transfer function rules}
\end{figure}

As we already defined the abstract domain and corresponding operations, the last step is to define the transfer function.
To this end, we define the transfer function that soundly approximates the concrete semantics.
The transfer function of intraprocedural program behavior analysis serves as the core of the analysis, and it is represented as a series of semantic rules.
$$
    \begin{array}{c}
        \Sigma;m;\Delta; v \Rightarrow e \\
        \hline
        \Sigma;m;\Delta, s \Downarrow \Sigma';m';\Delta'
    \end{array}
$$

The premise of the judgment is read as,
``with local variable context $\Sigma$, memory $m$, and pointer mapping container $\Delta$, varible $v$ evaluates to $e \in \mathcal{L}$.''
The conclusion of the judgment is read as,
``with local variable context $\Sigma$, memory $m$, and pointer mapping container $\Delta$,
statement $s$ produces a new local variable context $\Sigma'$, memory $m'$, and pointer mapping container $\Delta'$.''
The details of the semantic is shown in Fig.~\ref{fig:transfer-function}.
We divide the semantic rules into two parts: memory-related rules and non-memory-related rules,
and discuss them separately.

\ding{182} Rule~\ref{rule:binop} and \ref{rule:unop} are standard symbolic execution rules, we evaluate the expression on the current environment,
and substitute the target variable with a new element according to the operator.
Rule~\ref{rule:call} is similar to the previous two rules, except that we create a new symbol $f(e_1,\dots,e_n)$ to represent the return value of the function call.
Rule~\ref{rule:phi} is used to handle the phi node, which evaluates to a new phi node based on the values of the incoming variables.
Rule~\ref{rule:seq} is just a composition of two statements that transfer the state.

\ding{183} Rule~\ref{rule:assign}, \ref{rule:ref}, \ref{rule:load} and \ref{rule:store} are used to handle the memory related statements.
To make the semantic rule clear, we separate the transfer function into two parts: 1) points-to relation $\Delta$ transfer; 2) and local variable context $\Sigma$ and memory $m$ transfer.
Symbol $\Downarrow_{pa}$ denotes arbitrary pointer analysis transfer function.
Since there exist a lot of pointer analysis algorithms, we can use any existing algorithms to implement the memory-related part.
Fig.~\ref{fig:pointer-analysis} is an instance of the pointer analysis.
So we do not discuss the details of the pointer analysis in this paper.
Besides, the two parts can be implemented separately.
On the other side, pointer analysis is a time-consuming process, it is reasonable to use a flow-insensitive pointer analysis to calculate the points-to relation,
which may not need to be calculated at every program point.
In that case, the points-to relation is calculated as a pre-analysis, and there is no need to update the points-to relation in the semantic rule.

\ref{rule:assign} and \ref{rule:ref} are straightforward since they simply update the pointer mapping $\Delta$ based on the statement.
For load statement $v_1=\ast v_2$, we use the points-to relation to obtain the possible memory location of variable $v_2$, then join the values of these memory locations to get the value of $v_1$.
For the store statement $\ast v_1=v_2$, we also use the points-to relation to get the possible memory location of variable $v_1$, then store the value in those memory locations.
Similarly to strong and weak updates in pointer analysis, we use the join operator to update the memory value when there are multiple memory locations.

It is worth noting that the pointer analysis is not a trivial part of this analysis.
In the ideal case, we do not need an extra pointer analysis to manage the memory information; the points-to relation can be directly calculated from the possible values of the variables.
The problem arises when we encounter statements like $*a = b$ and $a\Rightarrow\top$ in practice.
The only sound way to interpret the statement is to assign to every possible memory location the value evaluated from $b$.
As a result, the entire analysis will be useless, since for all memory cells $o$ in $m$, $m(o)=\top$. Therefore, in store and load statement, we look up the points-to relation for the address variable from pointer analysis instead of evaluating the address.

As an intra-procedural analysis, we do not need to consider the effects of the callee function.
To be specific, we make two simplifications in the analysis. We assume:
\begin{itemize}
    \item The function call does not modify the memory pointed by the global variable and its arguments.
    \item The memory region pointed by parameters and global variables, directly or indirectly is disjoint.
\end{itemize}

The reader may doubt the soundness of the analysis after simplification.
In general, a function call $f(v_1,\dots,v_n)$ can modify the memory pointed by global variable and its arguments, which is ignored in our rule~\ref{rule:call}.
We argue that in intraprocedure, it is a reasonable simplification to ignore the side effects of the callee function.
It does not affect our analysis, since we are analyzing the behavior of the given function itself. All other function calls are treated as a black box.
We do not need to know the exact behavior of the callee function to analyze the function considered in the intraprocedural.
In the other hand, the whole analysis can be too imprecise to be useful if we consider the callee function and set all related memory cells to $\top$, thereby losing track of all information.
When we expand the analysis to interprocedural, \ref{rule:call} is replaced by its actual semantic of the callee function~\ref{rule:call2}.
The same reason is also applied to the second simplification.

To ensure the effectiveness of the worklist algorithm, we also need to guarantee the monotonicity of the proposed transfer function.
For statements that do not involve memory, the monotonicity is obvious based on the partial order rule in Fig.~\ref{fig:partial-order}.
Taking binary operation $a = b + c$ as an example, suppose $ b \Rightarrow l_1 \sqsubseteq l_1'$ and $c \Rightarrow l_2 \sqsubseteq l_2'$, then $a\Rightarrow$ binop($l_1,l_2$) $\sqsubseteq$ binop($l_1',l_2'$).
Suppose the points-to information $\Delta$ is monotonic, the transfer function of store statement is also monotonic since we just store the value to every related memory location.
Load statement is also monotonic due to the monotonicity of the join operator.

Suppose that the points-to-information $\Delta$ is calculated by a sound pointer analysis.
Program behavior analysis performs essentially a flow-sensitive analysis, the join of program state is defined from the join of the lattice.
Thus, program behavior analysis is sound if the pre-analysis is sound.
In summary, our analysis results are safe approximations of the actual behavior of the program.

Equipped with above techniques and language, we can give the following definition:
\begin{definition}[Possible Value Analysis]
    For a given function $f$, intraprocedural \textit{possible value analysis (pva)} aims to compute the abstract value $l \in \mathcal{L}$
    for every variable and memory cell at every program point, i.e. $\text{pva}: (\mathcal{V} \bigcup \mathcal{O}) \to \mathcal{L}$.
\end{definition}

\begin{theorem}[Soundness]
    Suppose $\alpha$ is the function that represents the transfer function and join, $\llbracket f \rrbracket$ is the concrete semantic of the single function, and $\llbracket f \rrbracket^*$ is the abstract semantic of the function,
    Then we have
    $$
        \alpha(\llbracket f \rrbracket) \sqsubseteq \llbracket f \rrbracket^*
    $$
\end{theorem}

Currently, our analysis is path-insensitive.
Our approach does not record the control flow information during the analysis like symbolic execution does, since we do not rely on the path constraint to solve or add certain assertions to the program.
It is also possible to add an assert statement to the program and constrain the behavior following the previous work~\cite{spa}.
Regardless of the path constraint, control flow information can be calculated since the condition itself is a lattice element.

\section{Discussion}\label{sec:discussion}

\subsection{Interprocedural Analysis}\label{sec:interprocedural-analysis}

We have discussed the intraprocedural analysis in the previous section, and the analysis can be expanded the analysis to interprocedural, i.e. from function level to program level.
Similarly to compare the behavior of two functions, suppose that we have two programs $P$ and $P'$, and we want to compare the behavior of the two programs.
From the perspective of black box, we do not care about internal function calls or memory modification since they are not observable objects.
In the most extreme case, all function calls are not recorded, and the whole program behavior can be reduced to calls to IO-related operations,
which corresponds to the common meaning of side effects exactly.

It is worth noting that the implementation of interprocedural program behavior analysis is simpler than the intraprocedural one due to less information required.
The semantic rule of the interprocedural analysis is shown in Fig.~\ref{fig:transfer-function-inter}. We mark the different parts of the rule with numbers.
Pointer analysis $\Downarrow_{pa}$ can be replaced with any global pointer analysis algorithm.
In interprocedural analysis, the modification to function parameters, variables, and every function return value can be directly returned to the function call point, rather than being recorded as a feature.
Therefore, in rule~\ref{rule:call2}, instead of recording the function call, we evaluate the callee function in the current memory context. The context of the local variable $\Sigma$ is represented by $\{v_i\mapsto e_i\}$, where $v_i$ is the parameter of the callee function and $e_i$ is the value of the parameter.
After the evaluation, we set the return value and the new memory context back to the caller function.
As common interprocedural analysis, we use the call graph to control the analysis process and the return value and context should merge when the function is called in different places.
The details of context sensitivity are omitted in the paper.

In practice, program is not only composed of user-defined functions, but also the standard library and third-party library (TPL) functions.
Our framework can easily handle library functions since we can just record them as done in the intraprocedural analysis.
The grain of the analysis can be controlled at any level by interacting using the interprocedural or intraprocedural rule.

\subsection{Limitation}
Many code changes still can be evaluated to $\top$ and are useless in practice.
Our approach can benefit from the improvement of various program analysis techniques;
we believe that the precision of the analysis can be further improved by more sophisticated pointer analysis algorithms or control flow analysis algorithms.
Benefiting from SSA form, our analysis is not bothered by multiple assignments to the same variable.
However, the SSA form in many implementations is partial such as LLVM~\cite{lattner2004llvm}, which means that the memory-related instruction is actually not ssa.
The precision of our approach can be limited by $e = \bigsqcup{m[o_i]}, \forall o_i \in \Delta(v_2)$ in~\ref{rule:load}.

\section{Related Work}\label{sec:related-work}

Program behavior analysis work is the main part of our work.
To the best of our knowledge, we are the first to propose a static analysis to represent the possible value of every variable and memory cell at each program point in a \textit{symbolic representation}, and we
used the representation to capture the semantic of code diff. 
Interval analysis~\cite{ai} is a classical static analysis technique to compute the possible value of a variable at each program point.
and value set analysis (VSA)~\cite{vsa} is a static binary analysis technique, 
which uses abstract interpretation to safely approximate the set of values of each data object at each program point.
The main difference between VSA and interval analysis and our work is that their concern is numerical value, 
so the analysis is on the numerical domain; 
while we consider the symbolic argument and memory, so the analysis is on a symbolic domain.

\section{Conclusion}\label{sec:conclusion}
In this paper, we propose a new type possible value analysis and design
a new symbolic lattice.
In this paper, we propose a new methodology using which we can capture the semantic of code diff by observing the behavior of the function.
To this end, we propose a new type of program analysis, program behavior analysis, to calculate possible values at each program point.
Furthermore, we extend the analysis to interprocedural and discuss possible future work and downstream tasks.

\bibliographystyle{ACM-Reference-Format}
\bibliography{ref}

\end{document}